# The SOFIA Telescope

Alfred Krabbe

German Aerospace Center, Institute of Space Sensors and Planetary Exploration,
Rutherfordstr. 2, D-12489 Berlin, Germany

**ABSTRACT**

The SOFIA telescope as the heart of the observatory is a major technological challenge. I present an overview on the astronomical and scientific requirements for such a big airborne observatory and demonstrate the impact of these requirements on the layout of SOFIA, in particular on the telescope design as it is now. Selected components of the telescope will be described in their context and functionality. The current status of the telescope is presented.

Keywords:

## INTRODUCTION

SOFIA[1], the <u>S</u>tratospheric <u>O</u>bservatory for <u>F</u>ar <u>I</u>nfrared <u>A</u>stronomy mounted onboard a Boeing 747SP will open a new era in MIR/FIR astronomy. Starting in 2002, SOFIA will offer German and American astronomers a unique platform, providing regular access to the entire MIR and FIR wavelength range between 5μm and 300μm part of which is otherwise inaccessible from the ground. SOFIA's 2.7m mirror and optimized telescope system combines the highest available spatial resolution with excellent sensitivity. SOFIA will operate in both celestial hemispheres for the next two decades. The predecessors of SOFIA, the Lear Jet and the KAO, have proven the concept of an aircraft observatory to be very flexible, long lasting, economic and technologically mature for most research areas in the IR. With SOFIA the dream of a big flying observatory is now becoming true.

In December 1996, NASA and DARA (now DLR) signed a memorandum of understanding (MoU) on the development and operation of SOFIA. DLR awarded a contract to MAN and Kayser-Threde to develop the SOFIA telescope, and NASA awarded a contract to the Universities Space Research Association (USRA) to develop the aircraft, the hangar with the science mission and control center (SSMOC) and to operate SOFIA. A recent overview over SOFIA can be found in Krabbe & Röser (1999) and Krabbe, Titz & Röser (1999). The basic characteristics of SOFIA are listed in Table 1. At this time (April 2000), SOFIA's telescope has successfully passed its critical design review (CDR), the most important step prior to actually building the hardware of the telescope. The following sections describe the current design and status of the SOFIA telescope and it's major components.

## THE OPTICAL LAYOUT

The optical layout of the SOFIA telescope is explained in Figure 1 and summarized in Table 3. The optical design of the telescope is a classical Cassegrain system with a parabolic primary and a hyperbolic secondary. The f-number of the primary, using the 2.5 m beam, is as short as 1.3 in order to fit the telescope into the aircraft. The system f-number of 19.7 corresponds to an image scale of 240 μm/″ in the focal plane. The secondary mirror is attached to a drive system providing focussing and optical alignment as well as a wide range of chopping capabilities, programmable by either a user supplied analogue or TTL curve or by the telescope control electronics. A flat tertiary mirror reflects the beam into the infrared Nasmyth focus, 300 mm behind the instrument flange. If the tertiary is replaced by a dichroic mirror, the transmitted optical light is reflected by a second tertiary 289.2 mm behind the dichroic and sent to the visible Nasmyth focus. There it feeds into a focal plane guiding camera system (FPI). Two on-axis imaging and guiding cameras are available: the wide field imager (WFI) and the fine field imager (FFI). Both of these cameras are attached to the front ring of the telescope. Details of their optical performances are summarized in Table 4.

---

[1] http://www.dlr.de/SOFIA, or http://www.sofia.usra.edu/

Some of the top-level requirements for the SOFIA telescope are listed in Table 2. Among those, the requirement for the optical performance of the telescope - 80% light circle diameter of 1.5″ - is rather moderate and relatively easily to achieve. For wavelengths shorter than 3 μm, the local seeing at 13 km altitude will be about 2″ to 4″. Beyond 15 μm, SOFIA will be diffraction limited (Table 3). The image stability requirement of 0.2″ during flight (Table 2) presents the major challenge of the telescope design. This does not come as a surprise because the telescope rests on a shaky ground and it is exposed to high-speed turbulent winds. The aeroacoustic noise in the open cavity as well as the residual vibrations of the aircraft during flight excite vibrations in the structure of the telescope. These vibrations are dominated by resonances with rather discrete eigenvalues in the range between 10 and about 110 Hz. Understanding those vibrations and keeping them under control will be the aim of extensive telescope testing during the first 2 years of operation.

**Table 1**  SOFIA Project Summary

| Project | |
|---|---|
| Operated by | USRA for NASA & DLR |
| Development began | January 1997 |
| First light | 20 |
| Operation | 20 years |
| Share | USA 80%, D 20% |
| Operation | |
| Number of flights | ~ 160 per year |
| Operating altitudes | 12 - 14 km, 39000 - 45000 ft |
| Nominal observing time | 6 - 9 hours |
| Crew | ~ 12: pilots, operators, technicians, scientists |
| Instrumentation | |
| Instrument changes | 15 - 25 per year |
| Research flights | ~150 per year |
| In-flight access to instrumentation | continuous |
| First light instruments | 10 + test camera |

**Table 2**  Telescope Top Level Requirements (Excerpt)

| | |
|---|---|
| Optical Layout | Generic Cassegrain, Nasmyth focus |
| Primary Mirror Ø: | 2.7m f/1.2 |
| Wavelength range: | UV - Radio  (0.3μm - 1600μm) |
| Temperature range: | -50°C ... +85°C |
| Pressure range: | 0 - 15 km altitude |
| Flights: | 4 per week, 960 h per year |
| Lifetime | 20 years |
| Weight: | ≤ 17t |
| Field of View | 8 arcmin +10 arcmin (chopping) |
| Image quality: | 80% energy within 1.5 arcseconds |
| Emissivity: | ≤ 15% |
| Image stability: | 0.2 arcseconds during flight |

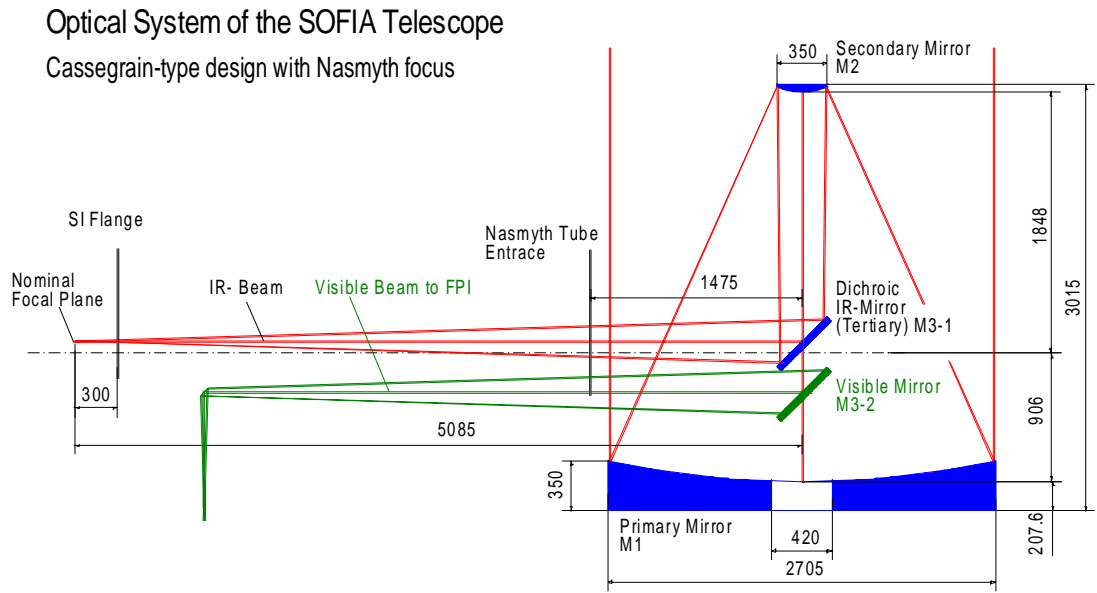

**Figure 1** The optical layout of the SOFIA telescope.

**Table 3** Optical parameters of the SOFIA telescope

| | |
|---|---|
| Entrance pupil diameter | 2500 mm |
| Nominal focal length | 49141 mm |
| Unvignetted field of view | ± 4 arcmin for chop amplitudes up to ± 5 arcmin |
| Aperture stop location | Secondary mirror |
| Aperture stop diameter | 352 mm |
| Primary free optical Ø | 2690 |
| "  focal length | 3200 |
| "  conic constant | -1 |
| "  center hole Ø | 420 mm |
| "  material | Zerodur (Schott) |
| "  mass | ~850 kg |
| Secondary radius | 954.13 mm |
| "  conic constant | -1.2980 |
| "  material | Silicon Carbide |
| e (Bahner, 1967) | 2754 mm |
| e+g (Bahner, 1967) | 6849 mm |
| Spatial resolution | 1-3″ for 0.3<λ<15μm |
| | λ/10″ for  λ > 15μm |

**Table 4** Optical data of the visual imaging and guiding cameras

| Parameter | Value |
|---|---|
| CCD chip size | 1024 x 1024 pixel |
| Pixel size | 14μm x 14μm |
| FFI  optical system | Schmidt-Cassegrain |
| primary diameter | 254 mm |
| night sensitivity | 13 mag (after 2.1 sec) |
| pointing stability | 0.35 arcsec |
| WFI optical system | Petzval |
| front lens diameter | 70 mm |
| night sensitivity | 8 mag (after 0.5 sec) |
| pointing stability | 1.8 arcsec |
| FPI  optical system | Eyepiece + commercial CONTAX 1.4/85mm lens |
| night sensitivity | 16 mag (after 1.7 sec) |
| pointing stability | 0.035 arcsec (nominal) |

## MAJOR COMPONENTS OF THE TELESCOPE

One of the design goals of the SOFIA observatory was to make every component as light as possible in order to maximize the useful observing flight duration. Every metric ton of weight saved translates into more than 8 minutes of additional observing time. Saving weight on the telescope itself is particularly important. All additional weight on e.g. the mirror support structure, counts fourfold: It is balanced by the counterweights on the opposite side of the telescope bearing and the telescope itself is balanced within the aircraft by counterweights in the front of the aircraft. Although the telescope as shown in

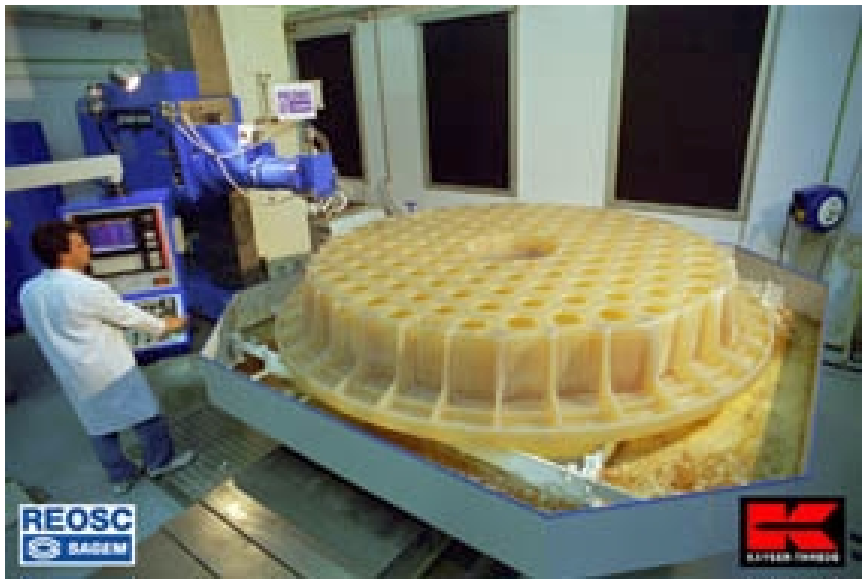

**Figure 2**  SOFIA's primary mirror as seen from the back side after lightweighting at REOSC, France.

Figure 5 only weighs about 17 metric tons, its structure still has to be very stiff. Therefore, most of the telescope structure, including the mirror support, the star frame, the truss work, the front ring and spider, and the Nasmyth tube, is made of carbon fiber reinforced plastic (CFC). The lowest eigenfrequencies of the telescope, the Dumbell bending modes, are just above 35 Hz.

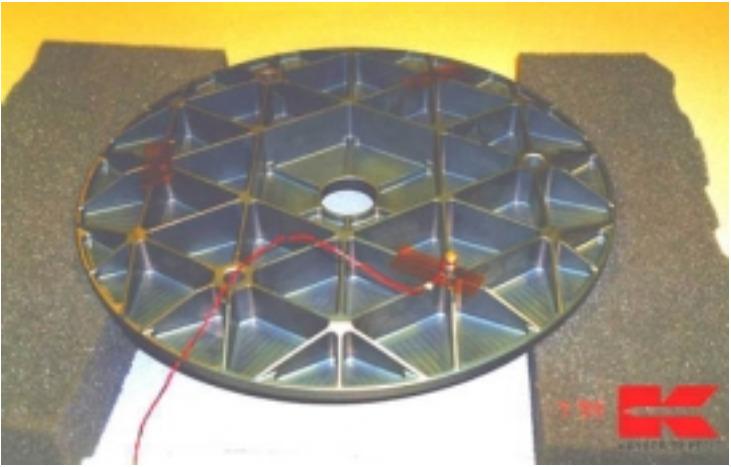

**Figure 3** The secondary mirror of the SOFIA telescope as seen from the backside. The secondary is made of silicon carbide (SiC), resulting in an extremely high eigenfrequency of 2 kHz at a total mass of 2.2 kg.

The primary mirror is made of Zerodur (Schott, Germany) and is currently being lightweighted and polished by REOSC, France (Figure 2). Delivery will be by the end of this year. Honeycomb shaped pockets have been milled into the backside of the mirror leading to a weight reduction by 80% corresponding to a final weight of about 870 kg. The wall thickness of the remaining hexagonal structure inside the pockets is only 7 mm. The central hole in the primary will take the tower, which carries both tertiary mirrors.

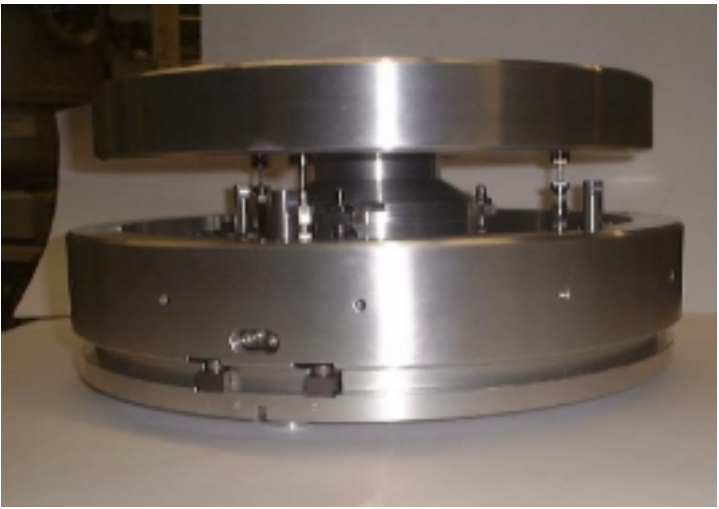

**Figure 4** Prototype of the upper stage of the SOFIA secondary drive system (CSEM, Swiss).

The secondary mirror, made of silicon carbide (SiC) has a diameter of 35 cm and hyperbolic geometry. The structure of the backside has been optimized for weight and momentum reduction, resulting in a wall thickness of 2 mm (Figure 3). Currently this is the largest secondary made of SiC. The hardness of the material in conjunction with the optimized design led to an free-free eigenfrequency of 2 kHz for the mirror itself at a weight of about 2.2 kg. The mirror will be completed by May 2000.

The secondary mirror is animated by a two-stage secondary drive system. A stiff hexapod driven base plate provides 6 degree of freedom movements for the secondary, which will be used for focussing and optical alignment as well as for optimi-zation of the image quality during operation. On top of the hexapod a 3-unit linear drive provides fast rotation movements of the mirror around its center of gravity. The linear drive system will be used for chopping, for user defined image motion as well as for image motion compensation of up to about 35 Hz. An interface plate provides stress-free acceleration of the mirror itself. The secondary mirror drive system has been designed as a very compact unit in order to fit into the limited space between the secondary and the cavity door, which is part of the aircraft. A lab prototype of the linear drive unit (Figure 4) has been tested very successfully.

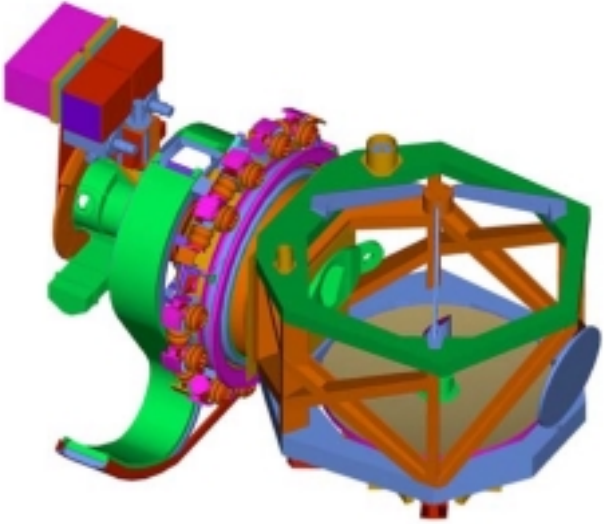

**Figure 5** Structural Assembly of the SOFIA telescope showing the location of the primary, the secondary, the tertiary tower, 2 guiding cameras on the front ring, the Nasmyth tube with the bearing, drive and vibration isolation system at the circumference, the cable eleviator and, at the outer left, the science instrument flange with counter weights and electronics boxes mounted above.

The bearing sphere embraces the Nasmyth tube and carries the dynamical weight of 9 tons of the telescope (Figure 6). Made of cast iron with an outer diameter of 1.2 m, it represents the inner part of the hydraulical oil bearing. The fitting of such bearings is rather tight since oil gaps around 40 µm have to be achieved. The bearing sphere is currently being polished to an accuracy better than 10 µm.

The coarse and a fine drive of the telescope provide a free elevation range relative to the aircraft of 20° – 60° during operation and 0° - 180° for maintenance. The telescope can be rotated in cross-elevation by ± 3° to compensate for aircraft movements. Rotation of the third axis by ± 3° provides a means of compensating image rotation during at least 12 to 15 minutes. The optional instrument rotator at the instrument flange will also provide image rotation compensation. Positioning of the SOFIA telescope is based on spherical sensors (position sensitive devices, PSDs), located at the bearing, and an high resolution inertial three laser gyro system, which is updated regularly by the focal plane tracker camera. The gyros were specially developed for SOFIA to provide extremely low drift rates and noise. Each of them consists of 3km of glass fiber providing an angular increment of 0.0008″.

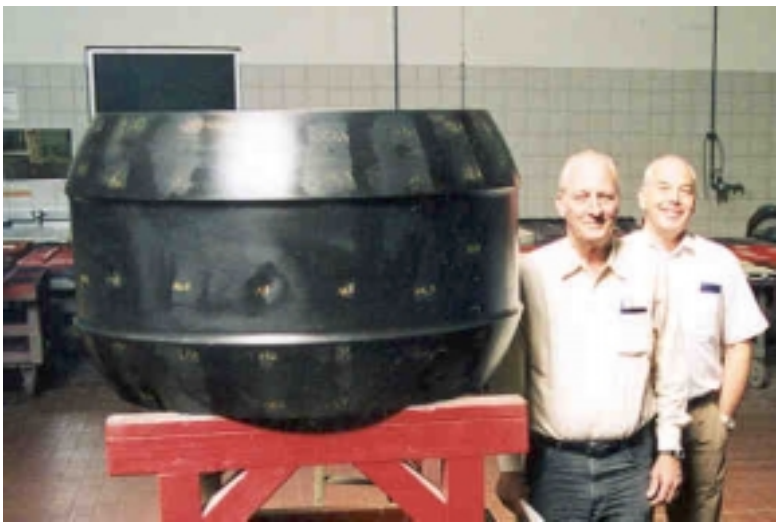

**Figure 6** Inner sphere of the oil bearing of the SOFIA telescope proudly presented by representatives of MAN, Germany. The final overall surface accuracy of 1.2 m bearing sphere will be < 10 µm.

The vibration isolation system consists of pressurized rubber bumpers, which are radially and axially arranged between the telescope drive and the forward bulkhead (Figure 5) and represent the structural interface between the telescope and the the aircraft. They can be adjusted to accommodate different altitudes and have been successfully tested (Figure 7).

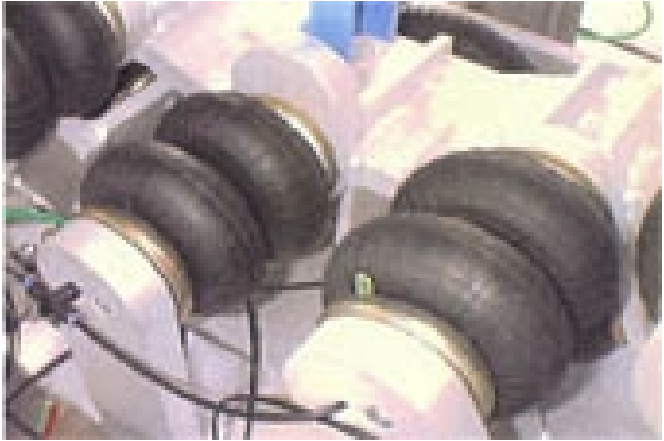

**Figure 7** Variable pressure can be applied to the rubber bumpers to accommodate the vibration isolation system for different altitudes.

During flight the optics and their supporting structure will be exposed to the environment at the altitude of 12 to 15km, while most of the Nasmyth tube, drive system and instrument flange are located within the cabin. Thus, the telescope has been designed to include an internal boundary with respect to pressure as well as temperature variations, both of which are substantial. The cabin will be precooled prior to each observing flight in order to maximize the precious observing time. After the observations before landing, the cavity will be flooded with dry nitrogen in order to prevent moisture from condensing on the optics and structure of the cold parts of the telescope.

According to the current schedule, the telescope will be integrated into the aircraft cavity during spring and summer of 2002 in Germany. First light is then expected for fall 2002. Regular astronomical observing will begin in December of 2002.